\documentclass{article}
\usepackage{authblk}
\usepackage{amsthm}
\usepackage{amssymb}
\usepackage{amsmath}
\usepackage{amsfonts}
\usepackage{wrapfig}
\usepackage{rotating}
\usepackage{pgfpages, multirow}
\usepackage{float}
\theoremstyle{plain}
\newtheorem{theorem}{Theorem}
\newtheorem{conclusion}{Conclusion}

\theoremstyle{definition}
\newtheorem{definition}[theorem]{Definition}

\begin{document}
\title{Multivariate time series approximation by multiple trajectories of a dynamical system. Applications to Internet traffic and COVID-19 data.}
\author[Victoria Rayskin]{Victoria Rayskin\\ victoria.rayskin@tufts.edu}

\date{\today}
\maketitle
\begin{abstract}
Utilization of multiple trajectories of a dynamical system model provides us with several benefits in approximation of time series. 
For short term predictions a high accuracy can be achieved via switches to new trajectory at any time.
Different long term trends (tendency to different stationary points) of the phase portrait characterize various scenarios of the process realization influenced by externalities. The dynamical system's phase portrait analysis helps to see if the equations properly describe the reality. We also extend the dynamical systems approach (discussed in~\cite{R5}) to the dynamical systems with external control.

We illustrate these ideas with the help of new examples of the rental properties HOMES.mil platform data. We also compare the qualitative properties of HOMES.mil and Wikipedia.org platforms' phase portraits and the corresponding differences of the two platforms' users. In our last example with COVID-19 data we discuss the high accuracy of the short term prediction of confirmed infection cases, recovery cases and death cases in various countries.
\end{abstract}

\section{Introduction}\label{intro}
The opportunities to obtain many different characteristics of data calls for the development of new methods for the analysis of multivariate time series. The traditional models (e.g., ARIMA, Vector Auto Regression, Holt-Winters) fit data into a single trajectory. The dynamical system models fit the data into a vector field, which provides infinitely many trajectories for various initial states. We can move from one trajectory to another for a better fitting. In this paper, we continue developing the dynamical system approach described in~\cite{R5} and introduce new examples of dynamical systems with control, new Internet platforms analysis and new COVID-19 data analysis.

It was shown in the groundbreaking works of Poincar\'e that many simple low order non-linearities in differential equations can produce complex (and even chaotic) trajectories. Using this idea, we want to construct simple low order differential equations that can accurately fit complex multivariate data. However, we want the dynamical system to be non-chaotic (adopting R. Devaney's definition of chaos). For example, we can consider systems that are monotone, or have reliably computable long-term tendency. Thus, our goal is to fit the data into the simple differential equations with {\it trending flow} (discussed in~\cite{R4}, see definition~\ref{def-trending}), which allow us to use the phase portrait for exploring the trends of the process. 
These trends can be associated with various scenarios of the realization of the processes or with the regime switches due to external effects at various times.
Moreover, the analysis of the dynamical systems' flow helps to understand if the model represents the real life sufficiently accurately.

As it was discussed in~\cite{R5}, the differential equations fitted to data allow us to obtain high precision (better than the VAR models) for the short-term prediction. In this paper we consider new examples which exhibit the same property.

The examples are based on the Internet platforms' data and on COVID-19 data. 

The volume of users interacting through the platforms is an important factor of the platforms efficiency. 
 The dynamical system approach allows us to study the future behavior and the tendency of the trajectories of the volume of each group of platform users. It also helps to increase the volume of users in the most cost efficient way with the help of the proper control of the dynamics. The example of the model with controlled dynamics shows which group of users should be incentified for the highest platform's benefits.

We illustrate these ideas using the data of the two platforms: HOMES.mil and Wikipedia.org. These platforms have different characteristics and qualitatively different phase portraits. It is a big advantage of the dynamical system models to reflect the ``physical'' properties of a process and to allow us to study these properties via phase portrait analysis.
The two different groups of users of the rental properties platform are interested in each other, but have a  within group competition (the property managers compete for the renters, while the renters compete with each other for being the most attractive candidate for the property managers). In the case of Wikipedia, the within group competition between Readers and Contributors is almost absent (with the exception of the ``edit wars'', which have a small effect of competition, discussed in~\cite{HKKR, HGMR}). 

In the example with the COVID-19 data we study the dynamics of the cases (confirmed infections, recoveries and deaths) of all significantly affected countries. In this preliminary study we minimize the total error of all countries' next day prediction. Even though the consideration of each individual country may produce even higher accuracy or have a similar accuracy for a longer forecast, our aggregate model still has a high precision.

The material is organized in the following way. In Section~\ref{section-error-analysis-2-dim} we compare the accuracy of the short-term prediction of the dynamical system and the Auto-Regressive models using HOMES.mil data. First, (Section~\ref{section-planar-model}) we do this for the two-dimensional case.
Then, (Section~\ref{sec-3D-model}) we construct the higher dimensional model and the model with control. In Section~\ref{sec-covid} we illustrate our dynamical system's approach with the examples of short term prediction of COVID-19 cases.

In Section~\ref{sec-phase-portrait} we discuss the phase portraits of the two- and higher-dimensional dynamical systems' model and the underlying characteristics of the process (long term trends). We also compare the qualitative properties of the dynamics of the model based on the HOMES.mil data and the model based on Wikipedia.org. We relate the difference in the dynamics to the difference in the users' behavior. 
\section{Description of the dynamical system model. Short-term prediction precision.}\label{section-error-analysis-2-dim}

In this section we describe the process of the dynamical system (DS) model construction, and then compare the precision of the short term forecasts via the DS model and via the traditional time series model. Among the traditional time series models, we choose the vector auto-regressive (VAR) model, which has good predictive power for multivariate time series. In each example, we tune the VAR model for fitting the data with the highest possible precision.

 Turns out that for the traffic data of rental properties platform (discussed below) and Wikipedia (see~\cite{R5}), the DS models perform better than VAR models. Our goal is to achieve the smallest short-term forecast errors.   The precision is measured as the sum of the squared distances between the true value taken from the testing data set and the value's short-term prediction, normalized by the sum of the squares of the testing values. Namely, we divide the data into two subsets, representing earlier time and later time. The first subset is used for the initial model construction. The second subset is used for the average error estimate. We optimize the sum of the squares of the errors, normalized by the sum of the squares of the testing values. 
 
When constructing the DS model, we assume that the noise does not affect the major law of the process, but shifts the realization of the process from one trajectory to another (corresponding to a new initial condition) of the major process. Starting the next time-step prediction from the position that precisely corresponds to the current state, allows us to apply the law of the process to the true (non-averaged) current state. This can help us to achieve a higher precision of the prediction.

In contrast, a traditional VAR (or similar) model can only use the current information for making more precise estimate of the average behavior.

Also, the precision of the DS prediction can be attributed to the fact that the model takes into account two conditions. The first one is the dependency between the earlier state and the later state  (derivatives of the model are calculated as the rate of change between two time-consecutive points). The second condition is the relation between the different coordinates of the multidimensional  state variable.

In this paper, the DS models related to Internet platforms are the systems of differential equations, which have polynomial right-hand sides. The different coordinates of the equations' variables correspond to the different types of platform's users or different attributes. 

The data for the rental property platform HOMES.mil consists of the monthly records between August 2017 and July 2019. This is a relatively small data set, which usually makes prediction challenging. 

We also compare the results of the HOMES.mil traffic model with the results of the Wikipedia's traffic model. The Wikipedia model is based on a bigger data set: the monthly 2008-2019 data set. The bigger data set helps to achieve a higher prediction's precision. Also, the interactions between the users of HOMES.mil and between the users of Wikipedia have different nature. For example, for the HOMES.mil, there is within group competition: buyers compete with each other for potential renters. For the Wikipedia, within group behavior is different. This distinction is reflected in the qualitatively different phase portraits of the two models. The Wikipedia model is discussed in greater detail in \cite{R5}.

Different trajectories of the dynamical system may be associated with different external conditions, different platform's policies and incentives that influence the system and force the transitions from one trajectory to another. These ideas may help platform owners to optimize their strategies.

The system of equations for COVID-19 data has a different form. Its right-hand side is approximated by the sum of the square root and polynomial functions. The different coordinates of the equations' variables correspond to the different countries' volume of cases. We create pairs of all significantly affected countries and derive two-dimensional DS models for all pairs. Our model allows to choose pairing that minimizes the error of the short term prediction of the volume of cases.

Each next day prediction can be initiated from a new trajectory that corresponds to the most appropriate current state of the process. These switches may be associated with different external effects, changes in the policy and in data accuracy. Our understanding of the properties of this dynamics may help us to develop the most effective strategies in our battle with the infection's spread.

\subsection{The two-dimensional models for the Internet platforms.}\label{section-planar-model}
First, let us consider the HOMES.mil platform.  We construct a DS model with the two variables: the number of Property Managers and the number of End Users (people searching for a rental property). For the model construction and the error estimate we use Property Managers and End Users monthly data of August 2017 through July 2019.

We search for the optimal parameters of the DS model, defined with the help of the equations~\eqref{seller-buyer-model} to model the traffic of Property Managers and End Users. The monthly number of users of each of the two types is denoted by the variables $x$ and $y$ correspondingly. The reason for choosing the equations of the form~\eqref{seller-buyer-model} (which generates the flow that we call {\it trending flow}) is discussed in the Section~\ref{sec-phase-portrait} and in greater depth in \cite{R3, R4, R5}.
\begin{equation}\label{seller-buyer-model}
\left\{
\begin{array}{l}
x' = \epsilon_1 x +V_1(y),\\
y' = \epsilon_2 y +V_2(x).
\end{array}
\right. 
\end{equation}

Comparing the errors of prediction for models having polynomial functions of degrees from 1 to 5, we find that the equations with the degree 3 polynomial functions of the form~\eqref{HOMES-eqns} give the smallest average error. We will call this model DS(3). 
\begin{equation}\label{HOMES-eqns}
DS(3):\ \left\{
\begin{array}{l}
x'=\epsilon_1 x +v_1y+v_2y^2+v_3y^3,\\
y'=\epsilon_2 y +w_1 x- w_2x^2+w_3x^3.
\end{array}
\right.
\end{equation}

Using the same method of the error estimate, we find that the best fitting autoregressive model is VAR(1). 

The error comparison (table~\ref{error-table-pure-2D-data}) shows that DS(3) model gives more precise prediction than VAR(1) model for the small data set of the rental platform. For a larger data set, such as Wikipedia traffic data, the DS model also gives a more accurate prediction than the VAR model (see table~\ref{wiki-error}, for details see~\cite{R5}). However, the difference is not as big as in the case of the very small data set. 

\begin{equation}\label{error-table-pure-2D-data}
\begin{tabular}{c||c|c|c}
{\multirow{2}{*}{Model}}&\multicolumn{3}{|c}{Average forecast error of the:}\\
 &Property Managers & End Users & Total\\
\hline\hline
DS(3)& .0023 & .0138 & .0161\\
VAR(1)& .0083 & .0177 & .0260
\end{tabular}
\end{equation}

\vspace{.4in}

\begin{equation}\label{wiki-error}
\begin{tabular}{c||c|c|c}
{\multirow{2}{*}{Model}}&\multicolumn{3}{|c}{Average forecast error of the:}\\
& Readers & Edits & Total\\
\hline\hline
DS(4)& .0051 & .0135 & .0186\\
VAR(2)& .0063 & .0134 & .0197
\end{tabular}
\end{equation}

\begin{sidewaysfigure}
\centering
\includegraphics[width=\textwidth,height=\textheight,keepaspectratio]{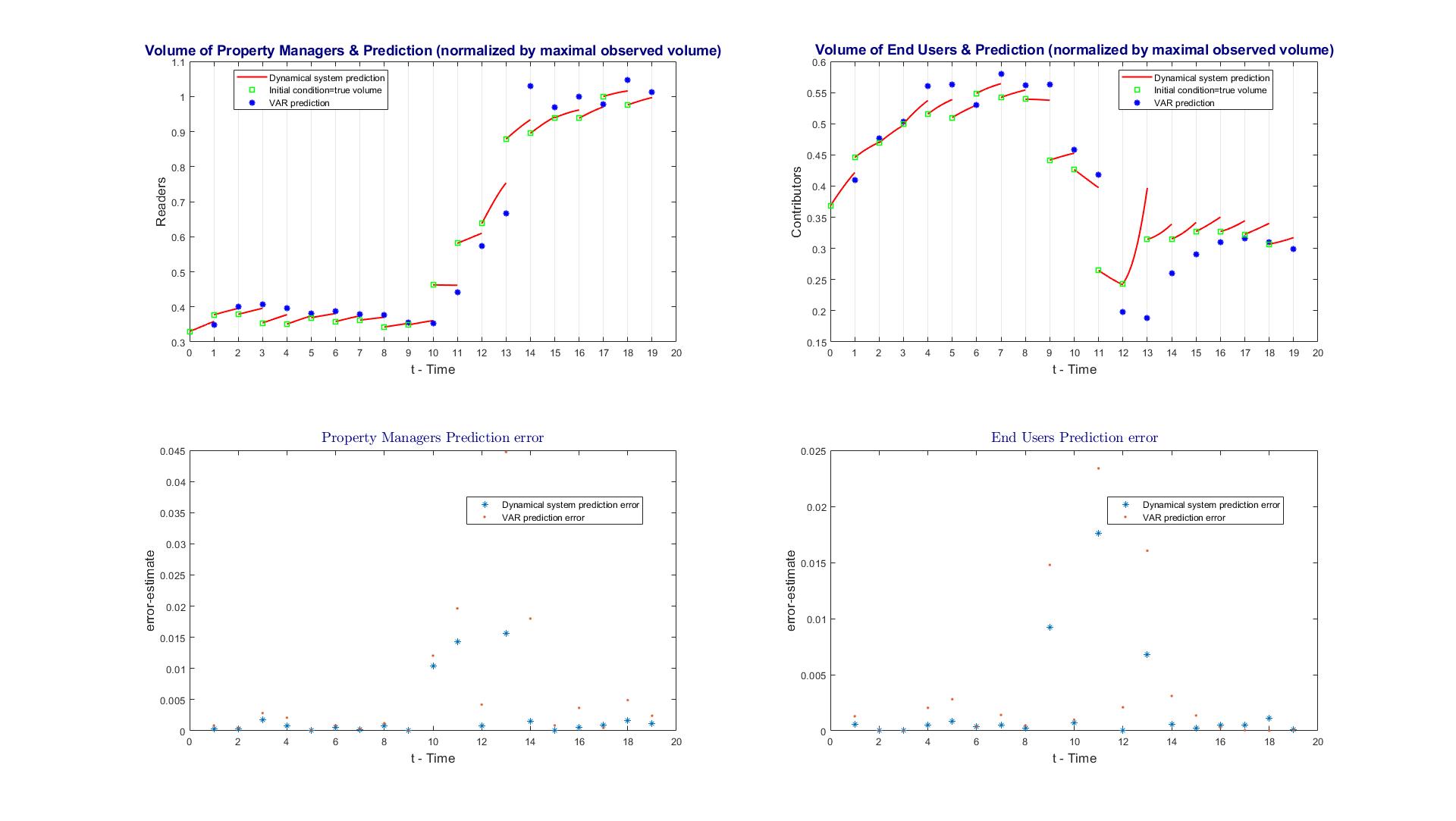}
\caption{The green squares in the upper Figures represent the volume of the Property Managers (left) and the volume of the End Users (right). The blue dots in the upper Figures are the VAR(1) estimates of the Volumes. The red arrows start at the initial condition (current state, shown as the green square) and end at the lag 1 predicted value. The lower Figures compare the errors of the DS(3) and VAR(1) models for each point from the testing data set. For the volume of Property Managers DS(3) makes a smaller average error than VAR(1). For the volume of End Users DS(3) makes smaller average error than VAR(1). I.e., on the average, the blue stars are lower than the orange dots.}
\label{Homes-2dim-prediction-and-error}
\end{sidewaysfigure}

The coefficients shown in~\eqref{HOMES-eqns-coefficients} were obtained for the model~\eqref{HOMES-eqns}, fitted in the entire  data set of the HOMES.mil platform. 
\begin{equation}\label{HOMES-eqns-coefficients}
\begin{array}{llllll}
\epsilon_1 & = & -0.1203, & \epsilon_2 &= & -0.2039\\
v_1 &= & 2.1307, & w_1 &= & 0.6754\\
v_2 & = & -7.6873, & w_2 &= & -1.4907\\
v_3 & = & 7.1497, & w_3 &= & 0.8933.
\end{array}
\end{equation}
The Figure~\ref{Homes-2dim-prediction-and-error} illustrates each testing data point prediction (upper figures) and error (lower figures), for each of the two variables. In the upper figures, the green squares correspond to the testing data points. These points are used as the initial conditions of the DS(3) model. The end of the red flow shows the predicted value for the next point in time. These predictions are more accurate (on average) than the VAR(1) predictions shown as blue dots in the upper figures. The lower figures show the error magnitude for each testing data point: blue stars for the DS(3) and orange dots for the VAR(1). On average, the blue stars are lower than the orange dots.

\subsection{The three-dimensional models for the Internet platforms.}\label{sec-3D-model}

Modern availability of various characteristics of the data calls for the development of the methods for high-dimensional data. Also, the interplay between the methods' order (complexity) and dimensionality of the methods provides better explanation of the real process that generates the data.

Our software allows us to create models of any finite dimension. Thus, for the rental platform's traffic  estimate we add one more variable: Referral Page Views. This variable corresponds to the number of all platform visitors (including Property Managers, End Users, personnel, unregistered potential renters and random platform visitors). Because the platform does not require End Users to register on the HOMES.mil platform for the property search, we cannot obtain accurate statistics of the potential renters searching for a property through the platform. Unfortunately, the 3 variables (the Property Managers, End Users and Referral Page Views) are not sufficiently independent characteristics of the platform users and in conjunction cannot benefit the predictive power of the model. 

For these 3 variables we have monthly recorded data for the period August 2017 - July 2019. The best fitting 3-dimensional dynamical system model is the linear DS(1) system of equations

\begin{equation}\label{HOMES-3D-model}
DS(1):\ \left\{
\begin{array}{l}
x_1'=\epsilon_1 x_1 + v_2 x_2 + v_3 x_3\\
x_2'=\epsilon_2 x_2 + w_1 x_1 +w_3 x_3\\
x_3' =\epsilon_3 x_3 + u_1 x_1 +u_2 x_2
\end{array}
\right.
\end{equation}

Here

\begin{equation}\label{HOMES-3d-eqns-coefficients}
\begin{array}{lllllllll}
\epsilon_1 & = & 0.0426, & \epsilon_2 &= & -0.0432, & \epsilon_3 &= & -0.3526\\
v_2 &= & -0.0033, & w_1 &= & -0.0139, & u_1 &= & 0.2971\\
v_3 & = & 0.0081, & w_3 &= & 0.0011, & u_2 &= & 0.1846.
\end{array}
\end{equation}

The origin is a hyperbolic fixed point (with the eigenvalues $-0.3473$, $0.0350$ and $-0.0409$). The volume of the users starts growing as soon as the initial number of users is sufficiently high.

\begin{sidewaysfigure}
\centering
\includegraphics[width=\textwidth,height=\textheight,keepaspectratio]{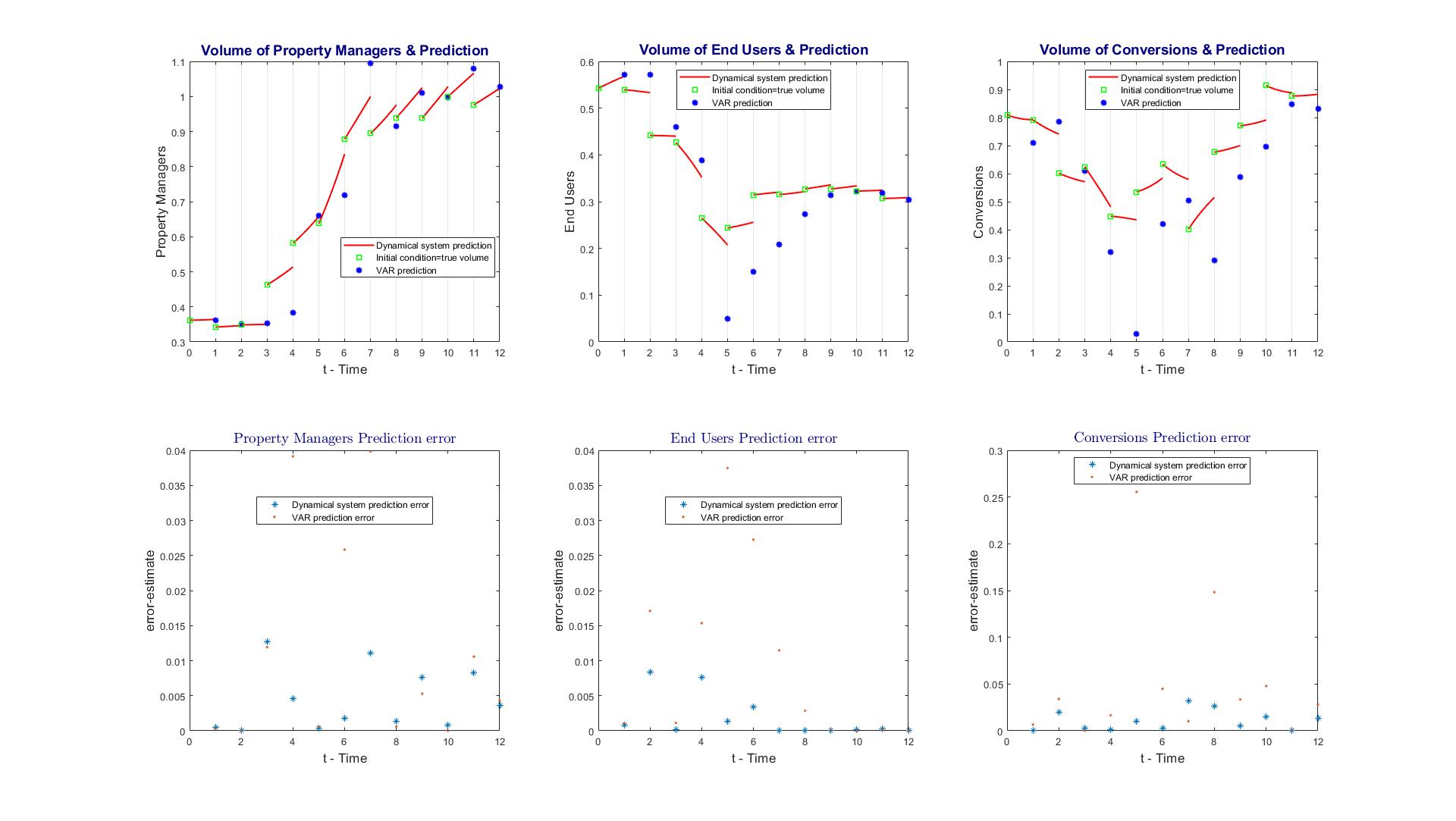}
\caption{The green dots in the upper Figures represent the volume of the Property Managers, End Users and Referal Page Views. The blue dots are the VAR(1) estimates of the Volumes. The red arrows start at the initial condition (current state, shown as the green dot) and end at the next time's value, predicted with DS(1). The lower Figures show that the DS(1) makes significantly smaller average error than VAR(1).}
\label{Homes-3dim-prediction-and-error}
\end{sidewaysfigure}

As it is discussed in \cite{BK} and \cite{GJB}, the analysis of complex time series data should benefit from the higher complexity models. The complexity can be viewed in several different ways. 
The higher order of VAR models can be reduced via introduction of new variables (if they are defined as back-lagged variables of the original VAR model). However, in the present example the data set is very small and the 3 variables are not sufficiently distinct. This is the reason, which makes the tree-dimensional model's error larger than the two-dimensional model's error. See Figure~\ref{Homes-3dim-prediction-and-error}.

Still, the average forecast error of the 3-dimensional best fitting DS model (DS(1)) is smaller than the average forecast error of best fitting VAR model VAR(1):

\begin{equation}
\begin{tabular}{c||c|c|c|c}\label{error-table-3D-data}
{\multirow{2}{*}{Model}}&\multicolumn{4}{|c}{Average forecast error of the:}\\
 &Property Managers& End Users & Ref Page Views & Total\\
\hline\hline
DS(1)&     0.0071  &  0.0149  &  0.0210 & 0.0430\\
VAR(1)&     0.0187  &  0.0759  &  0.1031 & 0.1977
\end{tabular}
\end{equation}

When we have a larger data set, higher dimensionality helps to increase the precision of some variables prediction. Returning to the Wikipedia example referenced in Section~\ref{section-planar-model}, in addition to the Readers and Edits we can consider the Contributors data. This three-dimensional model provides a more accurate estimate of Edits, as one can see, comparing the Table~\ref{wiki-error} and Table~\ref{error-table-3D-data-Wiki}. For details see~\cite{R5}.
\begin{equation}
\begin{tabular}{c||c|c|c|c}\label{error-table-3D-data-Wiki}
{\multirow{2}{*}{Model}}&\multicolumn{4}{|c}{Average forecast error of the:}\\
 &Readers& Edits & Contributors & Total error\\
\hline\hline
DS(4)&     0.0051  &  0.0116  &  0.0020 & 0.0187\\
VAR(3)&     0.0053  &  0.0132  &  0.0031 & 0.0216
\end{tabular}
\end{equation}

Even though the three-dimensional model for the HOMES.mil is not good, it is useful to consider higher-dimensional models for the controlled dynamical systems.  For example, the platform owners can control the flow of users with the help of incentives, represented by the input functions $u$ and $v$. More specifically, the original two-dimensional system~\eqref{seller-buyer-model} can be controlled in the following way:
\begin{equation}\label{control-model}
\left\{
\begin{array}{l}
x' = \epsilon_1 (x+u) +V_1(y+v),\\
y' = \epsilon_2 (y+v) +V_2(x+u),\\
u = k_1(t),\\
v=k_2(t). 
\end{array}
\right. 
\end{equation}
In our numeric examples we choose $k_1,\ k_2$ to be constant input functions. Obviously, they do not change the flow, but can be viewed as incentives that shift the initial conditions from one state to another. The output function $E$ (the third dimension shown on the Figure~\ref{figure-control}) measures the benefit of the platform. This benefit is higher if the number of End Users and  Property Managers is higher and it is reduced by the cost that the platform has to pay for providing the incentives to each group of users ($c_1,\ c_2$ correspondingly):
$$E(t) = p_1x(t)+p_2y(t) - c_1k_1(t) -c_2k_2(t).$$
Here $p_1, \ p_2$ represent the "benefit per user" that the platform receives  from each of the two types of the users. For definiteness, in our numeric example we assume that $p_1=p_2=c_1=c_2=1$. 

In our HOMES.mil example the controlled incentives shift the initial conditions to a new level and may help to move the dynamics into a higher basin of attraction, in which the volume of users can grow on its-own. 

The Figure~\ref{figure-control} compares 4 examples of incentives. The first case (upper-left) is when the platform provides no incentives ($u=v=0$). The benefit of the platform is the smallest for the majority of the simulated initial states of the system. The second case (upper-right) shows higher benefit for a slightly higher number of initial states. This is the case, when incentives are provided to the Property Managers only ($u=1,\ v=0$). The third case (lover-left) shows the best platform benefit, which corresponds to the incentives provided to the End Users ($u=0,\ v=1$). In this case the flow of users of both types starts growing on its-own and no additional incentives are needed. The last case (lower-right) corresponds to the incentives provided to both: Property Managers and End Users. The phase portrait shows that the flow of users of both types also grows on its-own. However, the cost of incentives provided to both groups of users is higher and makes the total platform benefit $E$ not as high as in the third case. 

\subsection{Model for the COVID-19 prediction of the volume of cases in multiple countries.}\label{sec-covid}
In this Section, we discuss our initial study of the COVID-19 data (published by John Hopkins University) with the help of the DC modeling approach. We focus on the short term (for simplicity, next day) prediction for all significantly affected countries (more than a hundred countries). For the accuracy of the short term prediction we drop the assumption that the DS model must have trending dynamics and allow the right-hand side of the equations to be expressed as the sum of square root and polynomial functions. Our goal is to minimize the total error of the predicted number of confirmed infection cases, recovery cases and death cases across all countries.

Our algorithm predicting the future number of cases consists of two parts. 

First, we create the country pairs. We use machine learning technique for dividing the set of all countries into pairs that would allow the most accurate prediction. Because the migration between geographically close countries affects the spread of the infection, one of the criteria of the pairing is the small distance between the paired countries. Two other characteristics of each pair is the similarity of the rate of growth and of the total number of the cases. 

\begin{figure}[h!]
\includegraphics[height=38mm]{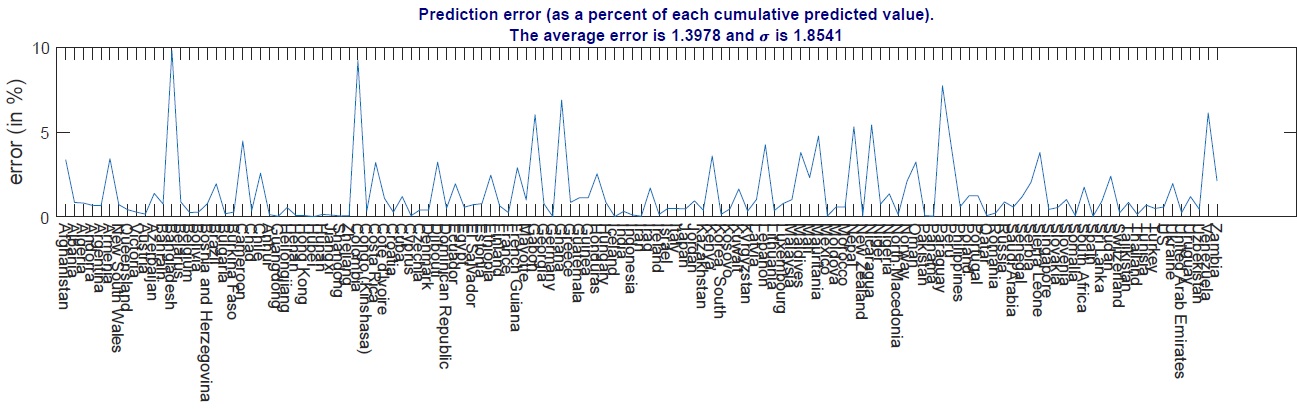}
\caption{Countries with more than 700 Recovery cases (as of 6/21) are included in the predictive model. The figure shows the prediction error of the Recovery cases by country, estimated by back-testing. The average error across all countries is $1.3978\%$.}
\label{recovery}
\end{figure}

\begin{figure}[h!]
\includegraphics[height=38mm]{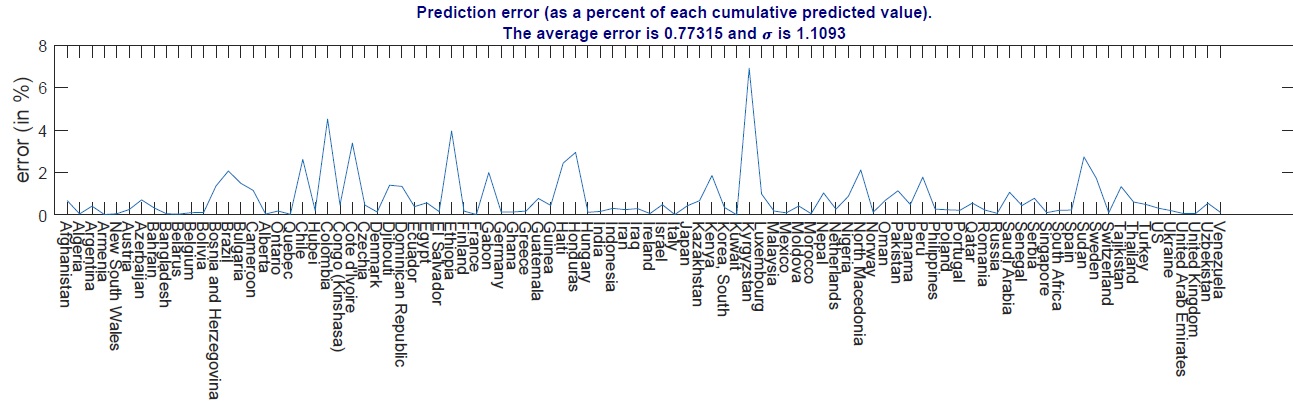}
\caption{Countries with more than 3000 Confirmed COVID-19 cases (as of 6/21) are included in the predictive model. The figure shows the prediction error of the Confirmed Infection cases by country, estimated by back-testing. The average error across all countries is $0.77315\%$.}
\label{covid}
\end{figure}

\begin{figure}[h!]
\includegraphics[height=38mm]{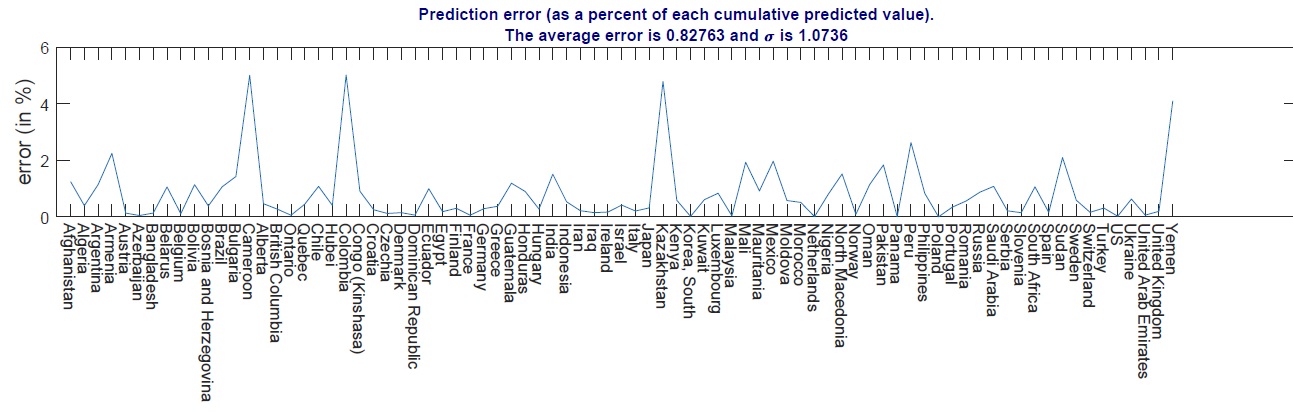}
\caption{Countries with more than 100 Death cases (as of 6/21) are included in the predictive model. The figure shows the prediction error of the Death cases by country, estimated by back-testing. The average error across all countries is $0.82763\%$.}
\label{death}
\end{figure}

Then, we apply the DS modeling ideas for the construction of the coupled differential equations. For each type of cases (confirmed infections, recoveries and deaths) we choose the type of model (\eqref{recovery-model}, \eqref{confirmed-model} and \eqref{death-model}) that minimized the total error across all countries, and after that we select the optimal country-specific coefficients\footnote{In these equations, $P_i$ and $R_i$ stand for the polynomial functions of degree $i$.}. The models accurately fit the data and allow us to make high precision short term predictions of confirmed cases, recoveries and deaths. The prediction errors are shown in Figures~\ref{recovery}, \ref{covid} and \ref{death}. 

Recovery cases:
\begin{equation}\label{recovery-model}
\left\{
\begin{array}{l}
x' = P_1(x,y)+a_1\sqrt{x}+a_2\sqrt{y}\\
y' =  R_1(x,y) + b_1\sqrt{x}+b_2\sqrt{y}
\end{array}
\right.
\end{equation}

Confirmed Cases:
\begin{equation}\label{confirmed-model}
\left\{
\begin{array}{l}
x' = P_2(x,y)+a_1\sqrt{x}+a_2\sqrt{y}\\
y' =  R_2(x,y) + b_1\sqrt{x}+b_2\sqrt{y}
\end{array}
\right.
\end{equation}

Death cases:
\begin{equation}\label{death-model}
\left\{
\begin{array}{l}
x' = P_3(x,y)+a_1\sqrt{x}+a_2\sqrt{y}\\
y' =  R_3(x,y) + b_1\sqrt{x}+b_2\sqrt{y}
\end{array}
\right.
\end{equation}
\begin{figure}[h!]
\begin{center}
\includegraphics[height=100mm]{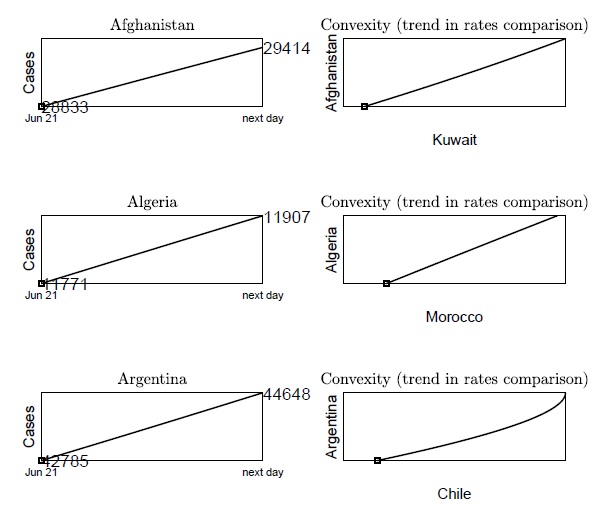}
\end{center}
\caption{This sample of the 3 countries from the alphabetical list of more than 100 countries shows the next day predictions and comparison of the rate of growth of infection in the neighboring countries.}
\label{sample}
\end{figure}

The Figure~\ref{sample} illustrates 3 of more than a hundred (listed alphabetically) country pairs' predictions that were tested with the algorithm for multiple days in April, May and June 2020.


\section{Trending flow and phase portrait analysis for the Internet platform models.}\label{sec-phase-portrait}

We have seen that the dynamical system models provide an accurate short-term forecast. In this Section we will provide new examples illustrating the idea that the dynamical system approach has another benefit. It allows us to see the main qualitative properties of the processes' dynamics. The global picture of the flow provides information about trends, which can be associated with various scenarios of the process's realization. Two- and higher-dimensional phase portraits also may help to detect other patterns and relations between the coordinates of the data.

VAR (or a similar traditional) model attempts to smooth-out the noise, and reflects the average process's behavior in a single trajectory. However, the true process may (depending on external conditions) have very different trajectories, and the realization of this process may oscillate between them, like in a well-known case of the two-dimensional dynamical system with a hyperbolic fixed point. In a small neighborhood of this point the process may oscillate between increasing and decreasing trajectories due to a small external distortion. In this case the trajectory, constructed via averaging of the data, does not provide us with a good explanation of the process. 
 
Stationary points are important objects in the phase portrait analysis. They are the landmarks that organize the long-term behavior and describe the major characteristics of a process. The types of stationary points explain the generic picture of the process and various scenarios of the process's realization. In our examples, we associate these scenarios with various basins of attraction of the fixed points. 

Phase portraits of some processes may have several basins of attraction. Fixed point in each basin shows tendency of the system in the long run. If the external conditions, regulations or incentives of the platform change significantly, the realization of the process may switch from one basin of attraction to another, and the two different fixed points can be associated with two different trends. Traditionally, this is modeled with the help of Intervention models, in which two different fixed points correspond to two different average behaviors. However, due to averaging, the Intervention models cannot carry as much information about the effects of the variation of the training data as the DS model. The DS models specifically reconstruct the effect of the rate of change of the training data.

Even though our software allows us to construct differential equations of arbitrary dimensions and with any right-hand-side functions, we do not want to use high complexity equations, which generate a chaotic flow, because simpler equations fit the data sufficiently well and may not necessarily be improved via higher complexity. Also, nonchaotic nature of the flow permits more reliable analysis of the global properties and trends of the process. In \cite{R4} we defined the {\it trending flow}, which we use here for the platform's traffic models.
\begin{definition}\label{def-trending}  
Assume that we are interested in the dynamics on the set $D$. Consider a system of differential equations defined on the domain $D$ (possibly, well-defined only on the interior of $D$). 
We will say that the system of differential equations has {\it trending flow on $D$} if its semiflow ($t\geq 0$) with any initial value in $D$ either converges to a fixed point in $D$, or escapes the domain $D$ (in finite time). 
\end{definition}

For the Internet platforms traffic, we use the system of differential equations (discussed in \cite{R4}) of the form~\eqref{seller-buyer-model-n-dim}: 
\begin{equation}\label{seller-buyer-model-n-dim}
\left\{
\begin{array}{l}
x_1' = \epsilon_1 x_1 +V_1(x_2,...,x_n),\\
x_2' = \epsilon_2 x_2 +V_2(x_1,x_3,...,x_n),\\
\vdots \\
x_n' = \epsilon_n x_n +V_n(x_1,...,x_{n-1}),
\end{array}
\right. 
\end{equation}

where for all $i=1,..., n$, $x_i\geq 0$ and $V_i\geq 0$  on the domain of interest $D$. 

If $n=2$, the flow is trending. Also, if $\epsilon_i \geq 0$ ($i=1,.., n$), the flow is trending. See~\cite{R4}. There are examples of trending flow in the class of monotone systems that has been studied in depth by M.W. Hirsch. See, for example, \cite{H1, H2, H3, H4, H5, HS} and references therein. 

The dynamics of these processes can be viewed with the help of the software and analysis presented in this paper.

\subsection{The planar phase portrait of the traffic of Property Managers and End Users on HOMES.mil.}\label{section-phase-portrait-2D-pure}

In this section, we discuss the phase portrait of the two-dimensional model~\eqref{HOMES-eqns} with coefficients~\eqref{HOMES-eqns-coefficients}, where the variables $(x,y)$ belong to  the domain $D=[0,\infty)^2$. 

We will also compare the qualitative properties of the planar phase portraits of the HOMES.mil  and Wikipedia platforms. The difference in the phase portraits properties can be associated with the distinct behavior of users of the two platforms.

\begin{figure}[h!]
\centering
\includegraphics[scale=0.19]{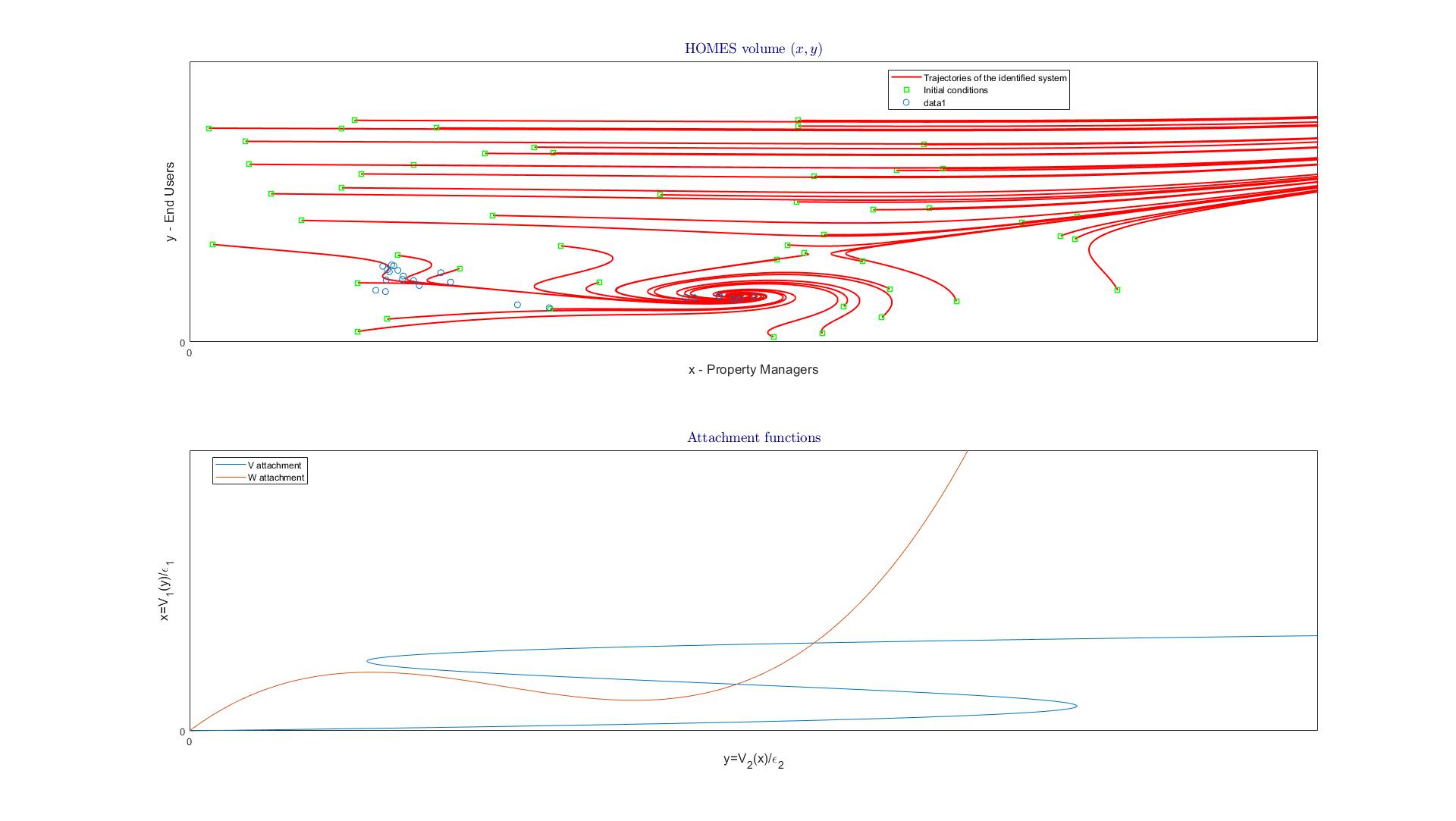}
\caption{The HOMES.mil platform starts growing with any low volume of users and tends towards the spiral attractor. If incentives are provided, the volume may progress into the second (higher) basin of attraction and  start growing to a much higher level.}
\label{HOMES_phase_portrait_pure_data}
\end{figure}

The HOMES.mil's flow in the domain of interest $D$ has three fixed points: two hyperbolic points (at the origin and in the interior of the first quadrant) and a spiraling attractor between them. The separatix is passing through the hyperbolic fixed point with positive coordinates. It separates (see Figure~\ref{HOMES_phase_portrait_pure_data}) the spiral sink's basin from the upper basin of attraction. Each of the fixed points is defined by the intercepts of the $x=-V_1(y)/\epsilon_1$ and $y=-V_2(x)/\epsilon_2$ graphs shown on the bottom of the Figure~\ref{HOMES_phase_portrait_pure_data}. 

In the system of equations~\eqref{HOMES-eqns}, $\epsilon_1, \epsilon_2$ are negative. The term $\epsilon_1 x$ is modeling the effect of the traffic of Property Mangers on their rate of growth, and $\epsilon_2 y$ is modeling the effect of the traffic of End Users on their rate of growth. These effects are negative because the users of the same type compete with each other (as discussed in \cite{R1, R2, R3, R4}). 
The values of $V_1(y), V_2(x)$ are positive on $D$ as shown on Figure~\ref{HOMES_phase_portrait_pure_data}.  $V_1(y)$ is showing how the volume of End Users affects the volume of Property Managers, and $V_2(x)$ is showing how the volume of Property Managers affects the volume of End Users, i.e. high volume of End Users attracts more Property Managers and high number of Property Managers stimulates more End Users to join the platform. 

The Wikipedia platform has a different behavior, that is captured by the model  defined as (see~\cite{R5}):

\begin{equation}\label{Wiki-eqns}
DS(4):\ \left\{
\begin{array}{l}
x'=\epsilon_1 x +v_1y+v_2y^2+v_3y^3+v_4y^4,\\
y'=\epsilon_2 y +w_1 x- w_2x^2+w_3x^3 +w_4x^4.
\end{array}
\right.
\end{equation}
with coefficients
\begin{equation}\label{Wiki-eqns-coefficients}
\begin{array}{llllll}
\epsilon_1 & = & -0.3570, & \epsilon_2 &= & -0.2243\\
v_1 &= & -0.2637, & w_1 &= &1.2710\\
v_2 & = &6.9566, & w_2 &= & -6.9038\\
v_3 & = & -16.4522, & w_3 &= & 13.6668\\
v_4 & = & 11.0347, & w_4 &= & -8.6907.
\end{array}
\end{equation}

The negative sign of the coefficients $\epsilon_1,\ \epsilon_2$ in~\eqref{Wiki-eqns-coefficients} can be explained by the ``edit wars'' (see~\cite{HKKR, HGMR}). However, as discussed in these papers, the negative effect is small, if compared with the trading platforms, where sellers compete with each other for buyers, and buyers prefer low volume of buyers, which makes them more attractive to sellers and assures lower prices. The Wikipedia's small negative effect creates small basin of attraction around the origin, and the platform owners need to provide small incentives when starting the platform, for the move into the basin of attraction of the positive fixed stationary point.

\begin{figure}[H]
\centering
\includegraphics[scale=0.2]{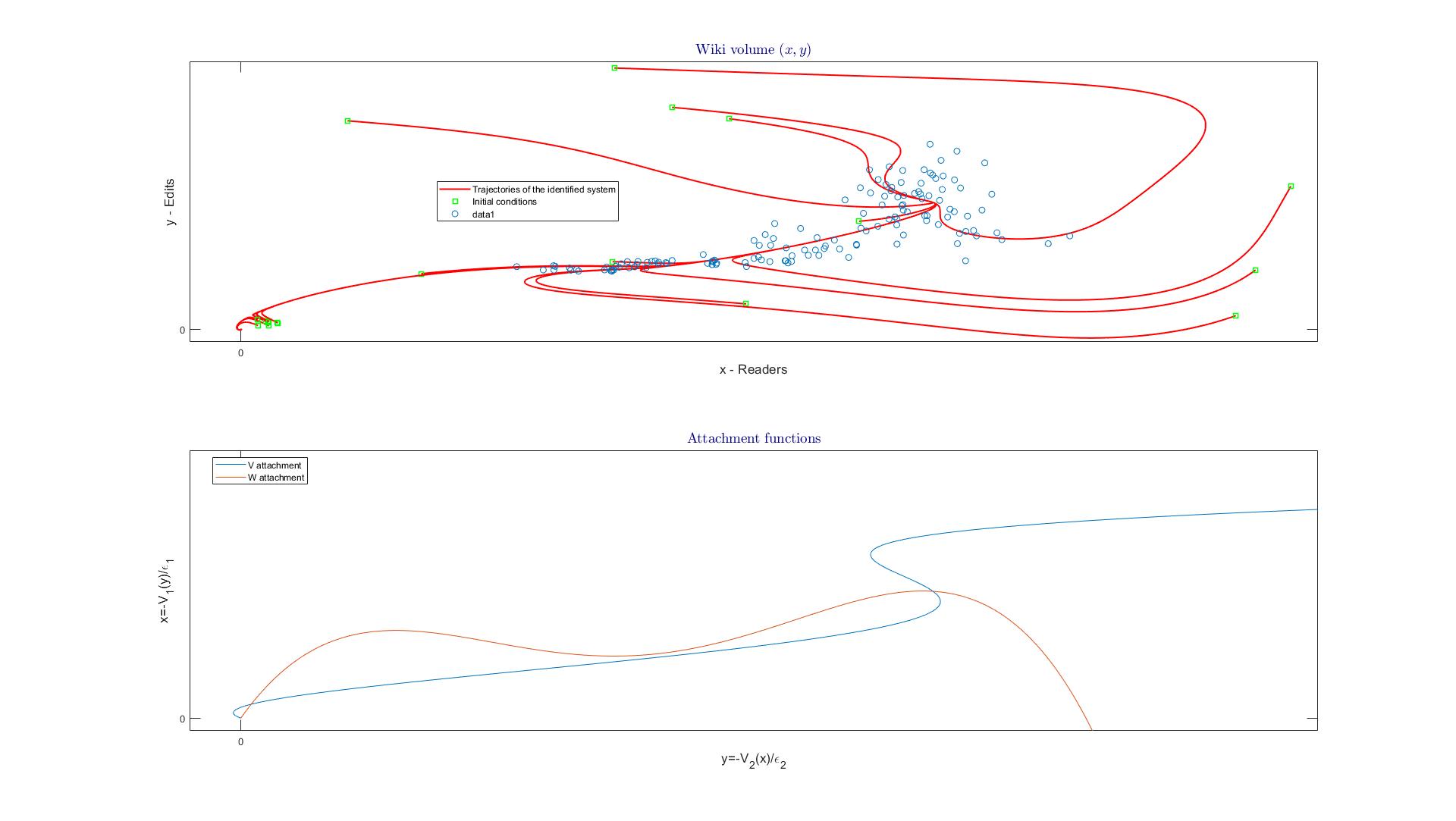}
\caption{The Wikipedia platform may start with initially low volume of users (in the origin's basin of attraction). If the volume  progresses into the second (higher) basin of attraction, then it  starts moving towards  the positive fixed point, which bounds the growth of users in the long run. If the volume gets above this fixed point (in one or both variables), it tends to eventually decrease towards the fixed point.}
\label{Wiki_phase_portrait_pure_data_near_separatrix}
\end{figure}

In the Wikipedia phase portrait's domain of interest $D$, there are three non-negative fixed points: the origin, the positive (close to the origin) point $a=(a_1, a_2)$ and the positive significantly larger fixed point $b=(b_1,b_2)$ ($0<a_1<<b_1$ and $0<a_2<<b_2$). The origin is the spiral attractor. The point $b$ is the attractor. The point $a$ is hyperbolic, through which the separatrix is passing. It separates (see Figure~\ref{Wiki_phase_portrait_pure_data_near_separatrix}) the origin's basin of attraction from the $b$ point's basin of attraction.

The function $V_1(y)$ is showing how the volume of Edits affects the traffic of Readers, and  $V_2(x)$ is showing how the volume of Readers affects the volume of Edits. $V_1(x)$ and $V_2(y)$ are positive on $D$ (see Figure~\ref{Wiki_phase_portrait_pure_data_near_separatrix} and \cite{R5}), i.e. high number of Edits attracts more Readers and high number of Readers stimulates more Edits.

It can be shown that both models (for the HOMES.mil and for Wikipedia platforms) generate trending dynamics (for the details please see~\cite{R2}, \cite{R3}). However, the characteristics of the platforms are different. Wikipedia platform after some initial incentives can start growing on its-own relatively early in its development (the lower basin of attraction is small), but in the long run it has a bounded growth of the volume of users, regardless the growth of the Internet users. The HOMES.mil platform tends towards its current state without any additional incentives, but if incentives are offered, it can switch into the higher basin of attraction and have significantly more users. As discussed in Section~\ref{sec-phase-portrait-3D}, it is optimal to offer incentives to the End Users.


\subsection{The three-dimensional model for prediction and control of the HOMES.mil dynamics.}\label{sec-phase-portrait-3D}

\begin{sidewaysfigure}
\centering
\includegraphics[width=\textwidth,height=\textheight,keepaspectratio]{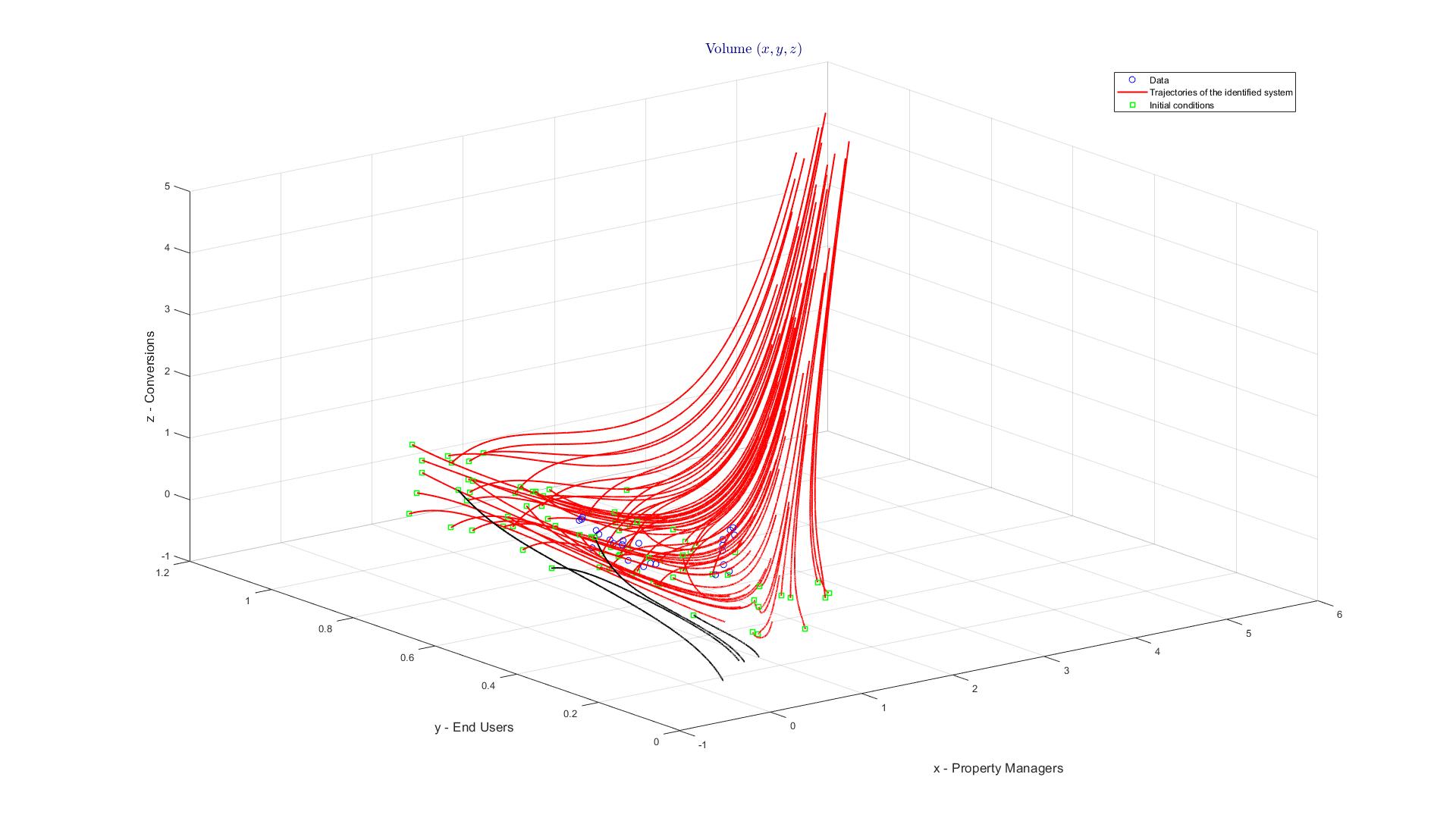}
\caption{The three-dimensional model does not fit the small data set of the HOMES.mil platform sufficiently well. The flow has non-realistic trends and confirms that this model should not be used for the platform's analysis.}
\label{Homes-3D-flow}
\end{sidewaysfigure}

In this section, we discuss the phase portrait of the tree-dimensional HOMES.mil model. We consider the system~\eqref{HOMES-3D-model} on the domain $D=[0,\infty)^3$. This is a simple linear model with hyperbolic fixed point at the origin. As discussed in Section~\ref{sec-3D-model}, its predictive performance is low. Moreover, as the Figure~\ref{Homes-3D-flow} shows its dynamics is not correct. For example, the black trajectories shows that the total number of Conversions approaches $0$ and escapes the positive domain, while the number of End Users and Property Managers are positive. This is not a realistic picture, which confirms that the three-dimensional model does not describe the platform's dynamics sufficiently well. We can only consider the trajectories above the separating hypersurface.  

\begin{sidewaysfigure}
\centering
\includegraphics[width=\textwidth,height=\textheight,keepaspectratio]{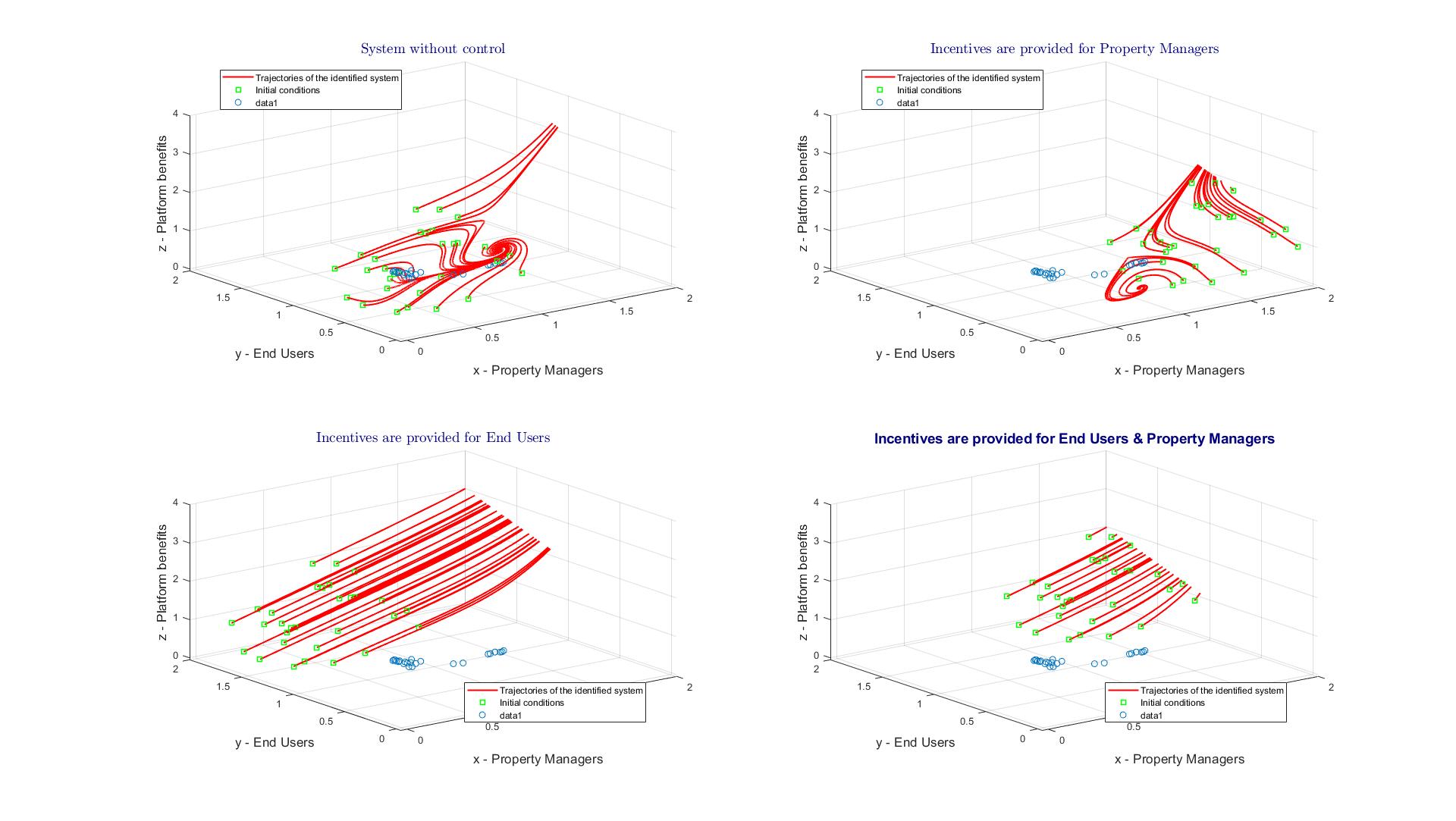}
\caption{Control variable adds incentives to Property Managers or End Users or both in the two-dimensional model. The highest platform benefits are achieved when the incentives are applied to the End Users.}
\label{figure-control}
\end{sidewaysfigure}

Even though the three-dimensional model for the HOMES.mil is not useful, the ability to visualize a three-dimensional picture can help to analyze the two-dimensional model~\eqref{control-model} with control. The model was discussed in the Section~\ref{sec-3D-model}. The coordinates $(E, \ x,\ y)$ help to see the level of platform's benefits corresponding to the varying number of users (see Figure~\ref{figure-control}).
\begin{conclusion}
{The phase portrait analysis of the HOMES.mil model shows that the highest platform's benefits may be achieved if the End Users are incentified. The level of incentives should be sufficiently high, so that the number of registered End Users is increased and the flow enters the higher basin of attraction. }
\end{conclusion}

\clearpage
\newpage
\begin{thebibliography}{KH}
%
\bibitem{BK} E. Bradley, H. Kantz, {\it Nonlinear time-series analysis revisited} (2015), Chaos, 25(9)
%
%
%
%
\bibitem{GJB} J. Garland, R. James,  E. Bradley, {\it Model-free quantification of time-series predictability} (2014) Phys. Rev. E {\bf 90(5)}. DOI: 10.1103/PhysRevE.90.052910
%
\bibitem{HKKR} A. Halfaker, A. Kittur, R. Kraut, J. Riedl, {\it A Jury of Your Peers: Quality, Experience and Ownership in Wikipedia} (2009). WikiSym Article 15, 10 pages. DOI:10.1145/1641309.1641332
%
\bibitem{HGMR}  A. Halfaker, R.S. Gieger, J. Morgan, J. Riedl {\it The Rise and Decline of an Open Collaboration System: How Wikipedia's reaction to sudden popularity is causing its decline} (2013) American Behavioral Scientist, {\bf 57(5)} 664--688, DOI:10.1177/0002764212469365 
%
\bibitem{H1} M. Hirsch, {\it Attractors for discrete–time monotone dynamical systems in strongly ordered spaces} Geometry and Topology: Lecture Notes in Mathematics 1167, 141--153. J. Alexander, J.Harer, editors. Springer-Verlag, New York, 1985.
%
\bibitem{H2} M. Hirsch, {\it Systems of differential equations which are competitive or cooperative. I: limit sets.} SIAM J. Appl. Math. 13 (1982), 167–179.
%
\bibitem{H3} M. Hirsch, {\it Systems of differential equations which are competitive or cooperative, II: convergence almost everywhere}, SIAM J. Math. Anal., 16, (1985) 423–439.
%
\bibitem{H4} M. Hirsch, {\it Systems of differential equations which are competitive or cooperative, III: competing species}, Nonlinearity 1, (1988) 51–71.
%
%
\bibitem{H5} M. Hirsch {\it On the nonchaotic nature of monotone dynamical systems}, European Journal of Pure and Applied Mathematics, {\bf 12}, No. 3 (2019), 680--688
%
\bibitem{HS} M. Hirsch \& H. Smith, Monotone Dynamical Systems, ``Handbook of Differential Equations,'' volume 2, chapter 4. A. Ca{n}ada, P. Dra{b}ek \& A. Fonda, editors. Elsevier North Holland 2005.
%
%
\bibitem{R1}V. Rayskin,  {\it Dynamics of two-sided markets, Review of Marketing Science}, {\bf 14}, (2016), 1--19.

\bibitem{R2}V. Rayskin,  {\it Users' dynamics on digital platforms}, {\em Mathematics and Computers in Simulation}, {\bf 142}, (2017), 82--97

\bibitem{R3}V. Rayskin,  {\it Users' Traffic on Two-Sided Internet Platforms. Qualitative Dynamics}, {\em American Institute of Physics Conference Proceedings}, {\bf 2164} 120012 (2019).

\bibitem{R4}V. Rayskin,  {\it Nonchaotic Models and Predictability of the Users' Volume Dynamics on Internet Platforms}, under review (2019)

\bibitem{R5}V. Rayskin, {\it Dynamical systems’ models for the prediction of
multi-variable time series. Wikipedia’s traffic}, preprint (2019)
%
%

\end{thebibliography}
\end{document}